\documentclass[10pt]{article}
\usepackage[OE]{express}
\usepackage{cite}
\usepackage[colorlinks=true,urlcolor=blue,linkcolor=black,citecolor=black,breaklinks=true,bookmarks=false]{hyperref}

\begin{document}

\title{Tunable erbium-doped microbubble laser fabricated by sol-gel coating}

\author{Yong Yang,$^{1,3,*}$ Fuchuan Lei,$^1$ Sho Kasumie,$^1$ Linhua Xu,$^2$ Jonathan Ward,$^1$ Lan Yang $^2$ and S\'ile Nic Chormaic$^1$}

\address{$^1$Light-Matter Interactions Unit, Okinawa Institute of Science and Technology Graduate University, Onna, Okinawa 904-0495, Japan\\
$^2$Department of Electrical and Systems Engineering, Washington University, St. Louis, Missouri 63130, USA\\
$^3$National Engineering Laboratory for Fiber Optics Sensing Technology, Wuhan University of Technology, Wuhan, 430070, China}

\email{$^*$yong.yang@oist.jp} %% email address is required



\begin{abstract}
In this work, we show that the application of a sol-gel coating  renders a microbubble whispering gallery resonator into an active device. During the fabrication of the resonator, a thin layer of erbium-doped sol-gel is applied to a tapered microcapillary, then a microbubble with a wall thickness of 1.3 $\mu$m is formed with the rare earth diffused into its walls. The doped microbubble is  pumped at 980 nm and lasing in the emission band of the Er$^{3+}$ ions with a wavelength of 1535 nm is observed. The laser wavelength can be tuned by aerostatic pressure tuning of the whispering gallery modes of the microbubble. Up to 240 pm tuning is observed with 2 bar of applied pressure. It is shown that the doped microbubble could be used as a compact, tunable laser source. The lasing microbubble can also be used to improve  sensing capabilities in optofluidic sensing applications.
\end{abstract}

 % REPLACE WITH CORRECT OCIS CODES FOR YOUR ARTICLE, MINIMUM OF TWO; Avoid using the OCIS codes for “General” or “General science” whenever possible.
%For a complete list of OCIS codes, visit: https://www.osapublishing.org/oe/submit/ocis/

%%%%%%%%%%%%%%%%%%%%%%% References %%%%%%%%%%%%%%%%%%%%%%%%%
%\bibliographystyle{osajnl}
%\bibliography{sensing}

%%%%%%%%%%%%%%%%%%%%%%%%%%  body  %%%%%%%%%%%%%%%%%%%%%%%%%%
%\textbf{Yong:1. P.Feron Paper is added, it is also temp tuning. If someone find more WGM tunable laser source let me know!\\
%2. As is broad band pumping, we don't know how much exactly power is coupled into the WGM. The 10 percent is an estimated amount of power lost after the MBR. So it cannot be used for threshold evaluation. Total power is more suitable.\\
%3. The wavelength of the 1550 nm laser is 1535.66, which is written in the next sentence.\\
%4. It is always a problem estimating the wall thickness accurately. However here it is more important that the laser is pressure tuned. We are not pursuing ultra thin wall microlaser this time, so I think this estimation is more than enough. Plus, pressure sensitivity is very sensitive to the wall thickness, from 1.2 to 1.4 um, the sensitivity changes from 9.2GHz/bar to 7.4GHz/bar. Updated.\\
%Fuchuan: 1.It's impossible to estimate the reflowing temperature and the size of the CO2 laser beam, so its power is also meaningless.\\
%}
\section{Introduction}
Whispering gallery mode resonators (WGR) are traveling wave resonators, which can have an ultra-high quality (Q) factors
and relatively small mode volumes. Due to these features, WGRs have been used for many recent research developments in areas as diverse as as cavity quantum electrodynamics (QED) \cite{Aoki2006}, optomechanics \cite{Aspelmeyer2014,Madugani2015a}, nonlinear optics \cite{Strekalov2016} and sensing \cite{Foreman2015,Ward2014}. Very high Q, low threshold WGR microlasers can be realized when the resonator is made from a material with gain \cite{Sandoghdar1996,Lissillour:01,Cai2004,Ward2010}. For this purpose, many methods have been developed \cite{He2013}, one of which is the sol-gel wet chemical synthesis technique. Rare earth ions are mixed into the sol-gel precursor solution and, based on the hydrolysis and condensation reactions of metal-alkoxide precursors in aqueous solutions, alcohol, or other media, a silica film can be formed with the gain medium diffused inside. Microlasers made from sol-gel coated microspheres \cite{Yang2003,conti2008} and microtoroids \cite{Yang2005} have already been realized. Such active WGRs are used for applications such as nanoparticle sensing \cite{He2011} and fundamental physics research\cite{Peng2014,Lei2014,LPOR201500163}. Sol-gel is a low-cost, flexible way to functionalize a WGR and can also be applied to microbubble resonators (MBR) as we shall discuss in the following.

MBRs are a more recently developed geometry of WGRs \cite{Sumetsky2010-mb,Watkins2011,Ward2014} that are hollow, while still maintaining a high Q-factor and small mode volume.  Similar to WGRs,  MBRs can also be used in a wide variety of applications,  such as nonlinear optics \cite{Li2013,Yang2016b}, sensing \cite{White2006,Henze2011,Yang2016}, and optomechanics\cite{Bahl2013}.  Active MBRs have also been developed by injecting dye solution\cite{Lee2011,Chen:16} or other bio-chemical liquids\cite{Fan2014} into the core of the resonator such that lasing emission can be achieved. In order to achieve lasing from the wall instead of the core, a glass-on-glass wetting technique was developed \cite{Ward2016}, whereby bulk Er-doped glass was melted onto the surface of a microcapillary. The wall thickness is limited in this method. To fabricate a thinner walled MBR with a gain activated wall and to improve the attainable sensitivity, alternative methods need to be found.

In this work, a sol-gel coating technique is used to introduce Er$^{3+}$ ions into the wall of an MBR and lasing in the 1550 nm band is realized. With a subwavelength wall thickness, the lasing is tuned by applying internal aerostatic pressure to the wall of the resonator.  Temperature\cite{Cai2004,Ward2010,Grivas2013} and mechanical\cite{Ta2013} tuning of laser sources based on WGRs have been reported previously. Refractive index tuning of the laser have been demonstrated\cite{Lee2011b} in the dye solution filled MBR. Here,the MBR laser is tuned alternatively by pressure without any gain medium in the core.

\section{Fabrication}
Having a high Q cavity mode is a necessary precondition to achieve low threshold lasing. For passive microbubbles, we have previously demonstrated that the Q-factor can reach $\rm 10^7$, which is close to the theoretical limit \cite{Yang2016}. To maintain the high-Q value after introducing a gain medium (erbium ions, in this case), we dissolved erbium ions into a sol-gel precursor solution and used this as the gain material, this ensures that the erbium ions are distributed uniformly in the silica matrix of microbubble after fabrication. The sol-gel precursor solution was made by mixing erbium(III) nitrate hydrate (i.e. $\rm Er(NO_3)_3\cdot 5H_2O$), tetraethoxysilane (i.e. TEOS), isopropyl alcohol (i.e. IPA), water ($\rm H_2O$), and hydrochloric acid ($\rm 37\%$ HCl) with a weight ratio of 0.03:6.5:6.1:0.7:0.6 for 2 hours at $\rm 70^{\circ}C$\cite{Yang2003}. After 24 hours, the sol-gel precursor was ready to use.

\begin{figure}[h]
\centering
\includegraphics[width=.8\textwidth]{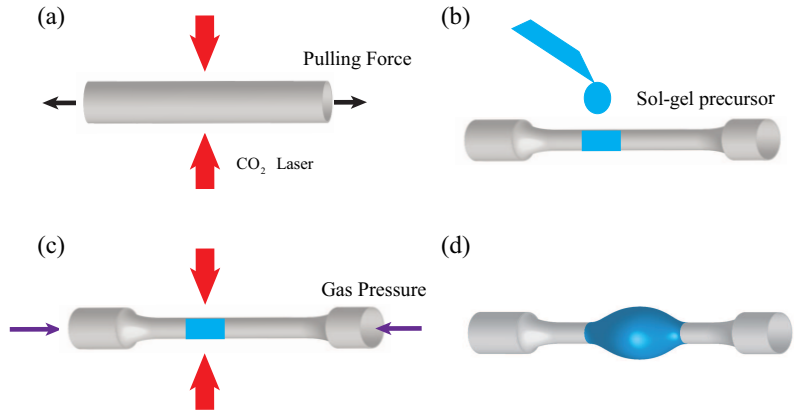}
\caption{Schematic of the fabrication process for a sol-gel coated microbubble using CO$_2$ laser heating. (a) A capillary is tapered using a $\rm CO_2$ laser heat source. (b)
Erbium ions are dissolved into a sol-gel
precursor, which is drop-coated onto the tapered
capillary. (c-d) The $\rm CO_2$ laser heats the sol-gel and internal air pressure is
applied until a microbubble is formed.}
\label{fig:fab}
\end{figure}

The fabrication of the erbium-doped microbubble is presented in Fig.\ref{fig:fab}. First, two counter-propagating $\rm CO_2$ laser beams were focused onto a silica capillary (outer diameter 350$\rm \mu m$, inner diameter 250$\rm \mu m$) to heat it, thereby allowing us to pull the capillary into a uniform taper with a waist diameter of around 30 $\mu$m. Afterwards, a droplet of the sol-gel precursor was transferred to the tapered capillary. Finally, the capillary was filled with compressed air and the $\rm CO_2$ laser was reapplied to reheat it. With the correct choice of laser power, the section of capillary in the focus of the laser beams expands to form a bubble. Due to the high temperature, the residual sol-gel solvent was removed and only silica doped with erbium ions remained; this material formed the wall of the bubble during the expansion process.The maximum erbium concentration was $\rm 5\times 10^{19}/cm^3$ according to its concentration in the sol-gel precursor.

\section{Aero-pressure tuning of the MBR laser}
\subsection{Experimental setup}
\begin{figure}[h]
\centering
\includegraphics[width=.5\textwidth]{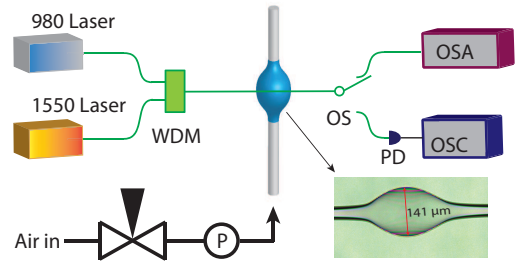}
\caption{Schematic of the setup for  pressure tunable lasing in an MBR. P: pressure gauge; OS: optical switch; PD: photo detector; OSC: oscilloscope; OSA: optical spectrum analyzer. The inset shows a  microscopic image of the sol-gel coated MBR. The diameter of the MBR is 141 $\mu$m.}
\label{fig:setup}
\end{figure}

The microbubble was coupled to a tapered optical fiber, using contact coupling. The laser source was a broadband, 980 nm laser with a maximum power of 200 mW. The tapered fiber waist was about 1.1 $\mu$m. Since the 980 pump has a width of about 2 nm, even without fine tuning of the laser frequency, some coupling into the MBR modes can occur. About 10$\%$ of the pump power was absorbed after passing the MBR. The coupled laser power excited 1550 nm lasing in the Er ions.  The experimental setup is illustrated in Fig. \ref{fig:setup}. In order to tune the frequency of the output laser, the MBR was placed in a dry nitrogen gas environment. Prior to this, one output of the MBR was sealed with epoxy, while the other end was connected to a compressed air cylinder. The internal pressure of the MBR was adjustable via a valve and the pressure reading was recorded on a pressure gauge connected to the gas source. During the experiment, the pressure was varied from 0 bar up to 3 bar. The lasing spectrum was measured on an optical spectrum analyzer (OSA), which had a minimum resolution of 0.07 nm.

\subsection{Experimental results}
\begin{figure}[h]
\centering
\includegraphics[width=0.65\textwidth]{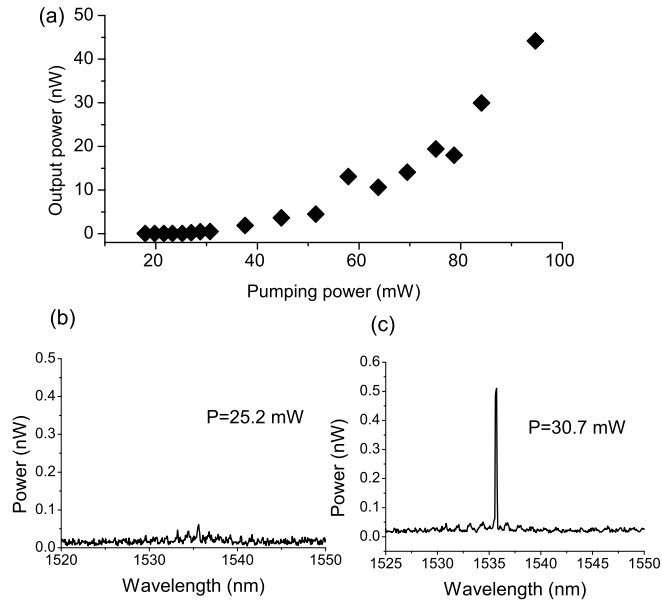}
\caption{(a) Lasing threshold measurement of the sol-gel coated MBR. The total power of the pump laser is used as an estimate. The threshold is about 27 mW for the pump laser. (b) The spectrum when the pump power is below threshold. (c) The single mode lasing spectrum near the threshold. }
\label{fig:threshold}
\end{figure}

First, the lasing threshold for the coated MBR was measured. The pump laser power was adjusted from 17.9 mW to 94.7 mW. Because the pump is broadband, it is difficult to know exactly how much power couples into the modes;  therefore, we used the total pump power in order to evaluate the threshold. The laser output power at 1535.66 nm was recorded against the pump power and is plotted in Fig. \ref{fig:threshold}(a). The plot exhibits typical laser output behavior as a function of the pump energy and the threshold is estimated to be about 27 mW. From the fluorescence background (see Fig. \ref{fig:threshold}(b)) single mode lasing occurs (see Fig. \ref{fig:threshold}(c)) when the pump power is beyond the threshold. In reality, the actual threshold should be significantly lower than this upper limit due to coupling losses.

\begin{figure}[htbp]
\centering
\includegraphics[width=.8\textwidth]{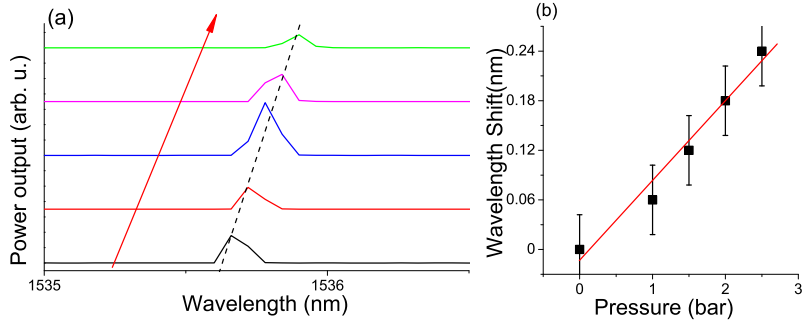}
\caption{(a) The laser spectrum of the sol-gel coated MBR at different pressures. The arrow shows the direction of the pressure increase. From bottom to top are the laser emission lines from 0 bar to 2.5 bar. The resolution of the spectrum is limited by the OSA. (b) The wavelength shift of the lasing mode as a function of applied pressure. The red line is a linear fit and the error bar is set by the resolution of the OSA.}
\label{fig:tuning}
\end{figure}

By applying aerostatic pressure inside the bubble, the MBR expands so that all modes are red-shifted \cite{Henze2011}. With a maximum applied pressure of 2.5 bar, the wavelength shift was about 240 pm (see Fig. \ref{fig:tuning}), which is much smaller than the bandwidth of the pump. Therefore, even without tuning the wavelength of the pump, modes can still be excited via the 980 nm source. The laser emission at 1535 nm is also shifted and this is shown in Fig. \ref{fig:tuning}(a). We have also plotted the wavelength shift as a function of applied pressure, see Fig. \ref{fig:tuning}(b). Note that this measurement is not so accurate since the laser linewidth and the shift rate are smaller than the resolution of the OSA. However, a linear relationship is still obvious, similar to the linear tuning property of the modes of a passive MBR \cite{Henze2011}.
\begin{figure}[h]
\centering
\includegraphics[width=1\textwidth]{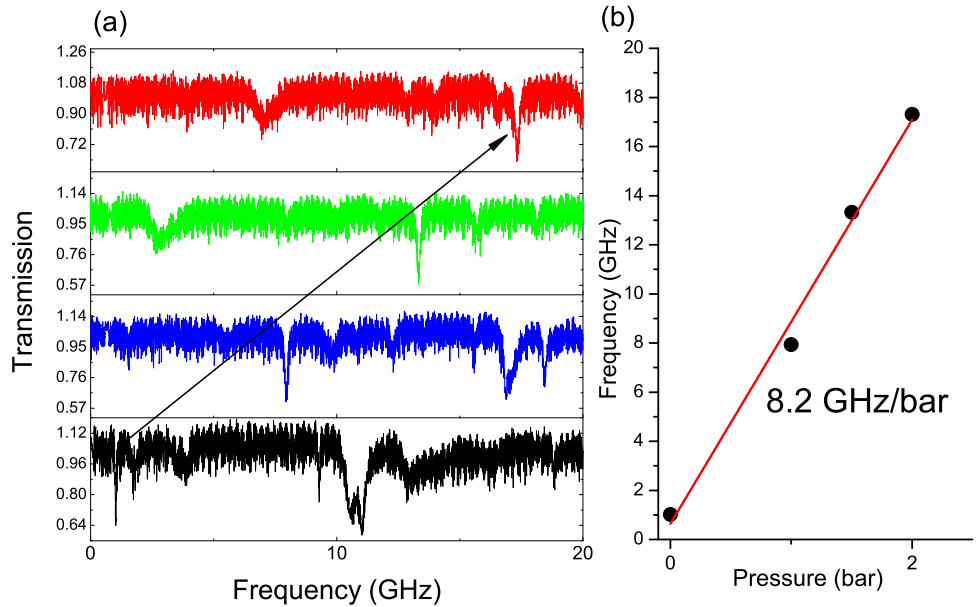}
\caption{(a) The transmission spectrum of the WGM at 1535.66 nm when scanned with 1550 nm laser source. From bottom to top are transmission spectra at 0, 1, 1.5, 2 bar respectively. The arrow shows the direction of the WGM shift due to the increase of the internal pressure. (b) Pressure tuning sensitivity fitted from the transmission spectra.}
\label{fig:trans}
\end{figure}

In order to obtain a more accurate pressure tuning slope, we measured the transmission spectrum by switching to a finely tunable laser, around 1535 nm (New focus Velocity 6728), and a photo detector connected to an oscilloscope, as shown also in Fig. \ref{fig:setup}). The transmission spectra of this laser, for different applied pressures are given in Fig. \ref{fig:trans}(a). The mode is indicated with an arrow in Fig.\ref{fig:trans} (a). The Q factor is above $10^7$ from the spectrum. The resolution of the pressure tuning, measured by monitoring the pump transmission through the fiber coupler, was limited only by the Q-factor \cite{Yang2016} and is, therefore, more accurate than the measurements made using the spectrum analyzer. By fitting to the resonance positions, as shown in Fig. \ref{fig:trans}(b), the tuning sensitivity was about 8.2 $\rm GHz/bar$. From the microscope image of the MBR, inset of Fig. \ref{fig:setup}, the wall thickness of the MBR was estimated to be about 1.3 $\mu$m. For an MBR with a diameter of 141 $\mu$m, the obtainable sensitivity is calculated to be 8.5 $\rm GHz/bar$\cite{Henze2011}, which is in accordance with the measured results.

\section{Conclusion}
By using a sol-gel technique, a layer of Er$^{3+}$ ions was coated onto the outer surface of a microbubble resonator. The Q-factor remained high for the MBR, which had an estimated wall thickness of of 1.3 $\mu$m.  We achieved lasing emission at 1535.66 nm when pumped at 980 nm. A pressure tunable laser was also demonstrated with a tuning range of 240 pm. This provides an alternative way of implementing a compact, WGR-based tunable laser source instead of the usual temperature tuning. A better tunable laser is expected since pressure tuning is linear and has reasonable long term stability \cite{Madugani2016}. Other rare earth ions can be diffused into the outer surface of the MBR instead of erbium through this way. So the wavelength of the MBR laser can be expanded to visible range, which falls into the transparency window for aquatic fillings in the core. The lasing modes of the active MBR then can improve the limit of detection for sensing applications\cite{He2011}.
\section*{Acknowledgment}
This work was supported by the Okinawa Institute of Science and Technology Graduate University.
\end{document}